# A (NOT SO) BRIEF HISTORY OF LUNAR DISTANCES: LUNAR LONGITUDE DETERMINATION AT SEA BEFORE THE CHRONOMETER

**Richard de Grijs**

*Department of Physics and Astronomy, Macquarie University,
Balaclava Road, Sydney, NSW 2109, Australia*
Email: richard.de-grijs@mq.edu.au

**Abstract:** Longitude determination at sea gained increasing commercial importance in the late Middle Ages, spawned by a commensurate increase in long-distance merchant shipping activity. Prior to the successful development of an accurate marine timepiece in the late-eighteenth century, marine navigators relied predominantly on the Moon for their time and longitude determinations. Lunar eclipses had been used for relative position determinations since Antiquity, but their rare occurrences precludes their routine use as reliable way markers. Measuring lunar distances, using the projected positions on the sky of the Moon and bright reference objects—the Sun or one or more bright stars—became the method of choice. It gained in profile and importance through the British Board of Longitude's endorsement in 1765 of the establishment of a *Nautical Almanac*. Numerous 'projectors' jumped onto the bandwagon, leading to a proliferation of lunar ephemeris tables. Chronometers became both more affordable and more commonplace by the mid-nineteenth century, signaling the beginning of the end for the lunar distance method as a means to determine one's longitude at sea.

## 1 THE MOON AS A RELIABLE GUIDE FOR NAVIGATION

As European nations increasingly ventured beyond their home waters from the late Middle Ages onwards, developing the means to determine one's position at sea, out of view of familiar shorelines, became an increasingly pressing problem. While latitude determination only required one to measure the height of the Sun above the horizon at its local meridian passage (i.e., at local noon), or that of Polaris, the North Star, at night, accurate longitude determination relies on knowing one's local time precisely with respect to that at a reference location. Prior to the development and general availability—and affordability—of accurate marine timepieces, careful observations of bright celestial objects were pursued as potentially viable alternative means to position determination.

In this article, I aim to provide a comprehensive review of historical applications to determine one's longitude at sea based on careful observations of the Moon. Perhaps surprisingly given these methods' importance in the history of maritime navigation, reviews of lunar longitude methods are often narrowly constrained in their focus, and so any scholar interested in obtaining a complete picture of the relevant developments must consult a wide range of sources, not all easily accessible. An exception is Cotter's (1968) *A History of Nautical Astronomy*, but that latter publication is no longer easily available. I will start by examining the use of lunar eclipses as reliable way markers, followed by a detailed exposition of so-called 'lunar distance methods' which became the most important shipboard techniques in use prior to the general adoption of chronometers.

The ancient Greek astronomer, Hipparchus of Nicaea (190–120 BCE; see Figure 1),[1] was convinced that accurate and internally consistent geographic maps should be based only on astronomical measurements of latitudes and longitudes, combined with triangulation. He adopted a zero meridian through Rhodes, in the eastern Mediterranean. Hipparchus suggested that one could determine positions East and West of his reference meridian by comparing the local time to the 'absolute' time determined at his zero meridian. He was the first to suggest that geographic longitude could be determined based on simultaneous observations of solar and lunar eclipses from geographically distinct locales:

> Many have testified to the amount of knowledge which this subject requires, and Hipparchus, in his Strictures on Eratosthenes, well observes, "*that no one can become really proficient in geography, either as a private individual or as a professor, without an acquaintance with astronomy, and a knowledge of eclipses. For instance, no one could tell whether Alexandria in Egypt were north or south of Babylon, nor yet the intervening distance, without observing the latitudes. Again, the only means we possess of becoming acquainted with the longitudes of different places is afforded by the eclipses of the sun and moon*". Such are the very words of

Hipparchus. (Strabo, 7 BCE: Book 1, Ch. 1)

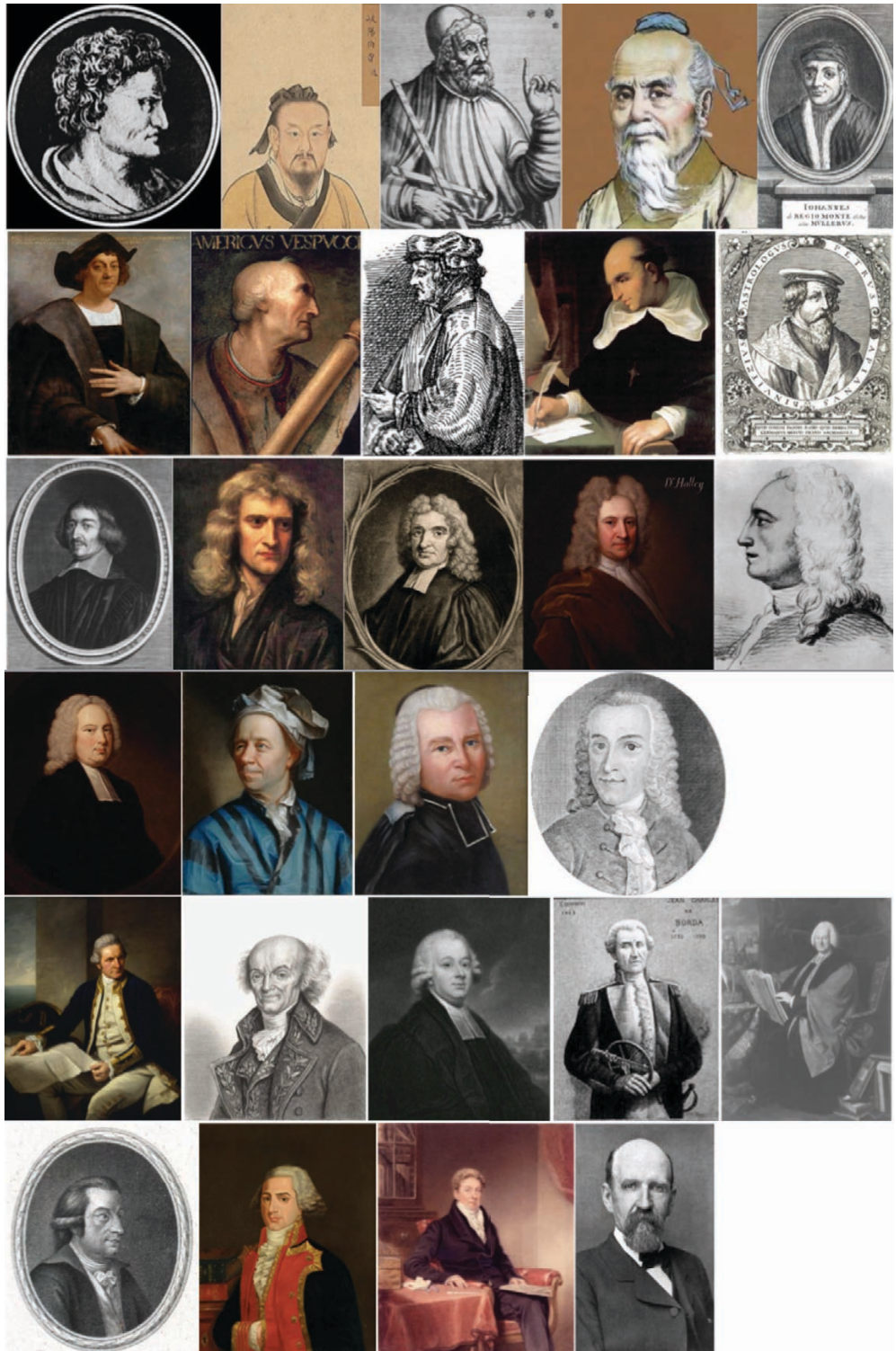

**Figure 1**: Portraits of the main characters driving the developments described in this paper, ordered from left to right and from top to bottom by date or birth. Individuals depicted include *(i)* Hipparchus of Nicea (190–120 BCE); *(ii)* Jia Kui (30–101 CE); *(iii)* Ptolemy (c. 100–c. 170 CE); *(iv)* Liu Hong (129–210 CE); *(v)* Regiomontanus (1436–1476); *(vi)* Columbus (c. 1451–1506); *(vii)* Vespucci (1454–1512); *(viii)* Werner (1468–1522); *(ix)* de las Casas (c. 1484–1566); *(x)* Apianus (1495–1552); *(xi)* Morin (1583–1656); *(xii)* Newton (1642–1726/7); *(xiii)* Flamsteed (1646–1719); *(xiv)* Halley (1656–1741); *(xv)* Hadley (1682–1744); *(xvi)* Bradley (1693–1762); *(xvii)* Euler (1707–1783); *(xviii)* de Lacaille (1713–1762); *(xix)* Mayer (1723–1762); *(xx)* Cook (1728–1779); *(xxi)* de Lalande (1732–1807); *(xxii)* Maskelyne (1732–1811); *(xxiii)* de Borda (1733-1799); *(xxiv)* Lyons (1739–1775); *(xxv)* von Zach (1754–1832); *(xxvi)* de Mendoza y Ríos (1763–1816); *(xxvii)* Norie (1772–1843); *(xxviii)* Slocum (1844–1909).

*Figure credits*: Wikimedia Commons (public domain), except for *(i)* http://www.ph.surrey.ac.uk/astrophysics/files/hipparchus.gif; *(ii)* https://www.epochtimes.com/gb/19/9/20/n11535183.htm; *(iv)* https://xw.qq.com/amphtml/20181109A1R8ZX00; *(xiii)* Creative Commons Attribution 4.0 International license; *(xiv)* National Portrait Gallery, NPG 4393; *(xvi)* NPG 1073; *(xx)* National Maritime Museum, BHC2628; *(xxi)* Joconde database entry 50350212050 (public domain); *(xxiv)* Cambridge University Library; *(xxvi)* Museo Naval de Madrid, Creative Commons CC0 1.0 Universal Public Domain Dedication; *(xxvii)* NPG 1131; *(xxviii)* U.S. Library of Congress, Prints and Photographs division, digital ID cph.3b46344 (out of copyright).

In fact, Hipparchus was the first to realise that one's longitude can potentially be determined by means of accurate time keeping (de Grijs, 2017: Ch. 2).

Meanwhile, halfway around the world, Chinese astronomers were already skilled in using lunar eclipses to determine their local longitude (Menzies, 2012). Their procedure involved a rather cumbersome process, however. It required synchronous time measurements across long distances and a base observatory in China. Observers at both the base observatory and the location of interest identified unique bright stars that passed through the local meridian at the moment the Moon started to reappear after an eclipse. Upon the travelling astronomers' return to their base observatory, both observation records were compared. A second measurement was obtained to determine the exact time difference of the meridian passages of the stars selected at each observing location. The measured time difference corresponded directly to the difference in longitude between both localities.

In addition, Chinese astronomers have long known about the Moon's irregular motion. The astronomer Jia Kui[2] (30–101 CE) was the first to address the Moon's irregular motion in the context of calendar design. In early records from 206 CE, Liu Hong used a linear interpolation method in his *Qianxiang li* to describe the Moon's daily motion (Qu, 2000). Liu's method retained the Moon's mean daily motion although adjusted for the orbit's intraday deviations. The Moon's elliptical orbit around the Earth causes it to travel faster across the sky at perigee (when the Moon is nearest the Earth) and slower at apogee.

The Moon's apparent motion takes it 360 degrees around the Earth to its original projected position among the 'fixed' background stars in approximately 27.3 solar days. This corresponds to an average of 13 degrees per day, or slightly more than half a degree per hour. While the fixed stars appear to move westwards because of the Earth's rotation, the Moon appears to describe a retrograde orbit in the eastward direction. The Moon appears to speed up when going towards the Sun and slow down when retreating. Its change in apparent motion also depends on the Earth's orbital position. The Moon's tilted orbit with respect to that of the Earth, inclined under an angle of 5.15°, causes it to perceptibly nod back and forth along its path.

Since Liu Hong's first attempt at linear orbital interpolation, Chinese astronomers have used a range of interpolation methods to predict the Moon's 'equation of centre'. Early accurate predictions of the lunar motion across the sky were obtained through the piecewise parabolic interpolation proposed by Liu Zhuo (544–610 CE; Qu, 2000). This method was further improved, using third-degree interpolation (although for the irregular *solar* motion), in compiling the *Shoushi* (1280) calendar by Guo Shoujing (1231–1316; Li and Du, 1987), the Chinese astronomer, engineer and mathematician whom the Jesuit astronomer Johann Adam Schall von Bell (1591–1666) called the 'Tycho Brahe of China'.

By simply measuring the angle on the sky between the Moon when it crosses the local meridian and a given star, one can calculate the local longitude. This requires only access to a sextant and a telescopic viewing instrument with a graduated arc to measure angles between specific points of observation, without the need for an accurate clock. This thus implies that Chinese astronomers would have been able to accurately calculate lunar positions on the sky since the Yuan Dynasty (1271–1368). In turn, this would have allowed them to compile ephemeris tables listing the projected positions of the Moon and a set of carefully selected bright reference stars throughout the year. This approach is, in fact, equivalent to the lunar distance method employed by European astronomers several centuries later. The method involved angular measurements on the sky of the distances between the Moon and a range of celestial objects (see Section 2).

While medieval Chinese astronomers had thus made significant progress in understanding the detailed lay of their land based on celestial observations, much of Europe was languishing in an environment dominated by religious intolerance and wars. Progress in science, technology and innovation was largely stifled.

On 3 August 1492, Christopher Columbus embarked on his first transatlantic voyage to discover 'Islands and Mainlands', with royal patronage. Columbus was familiar with Claudius Ptolemy's (c. 100–c. 170 CE) work, particularly with his description of what may be the earliest record of a difference in longitude. On 20 September 331 BCE, Alexander the Great's Macedonian army had witnessed a widely reported total lunar eclipse at its encampment near Gaugamela/Arbela (present-day Erbil, northern Iraq; Westfall and Sheehan, 2015). Ptolemy recorded that the eclipse was seen at the fifth hour of the night. In Carthage (near present-day Tunis, Tunisia), however, the eclipse occurred at the second hour of the night. This timing difference thus implied a distance between Arbela and Carthage of 45 degrees (Evans, 1998: 51), that is, 3 hours times the Earth's rotation rate of 15 degrees per hour. The actual longitude difference between both localities is, however, only 33 degrees. This large error is most likely attributable to the lack of information on *when* during the eclipse it was observed in either location (at the onset, at mid-totality?), as well as to difficulties related to the precision of local timekeeping (Westfall and Sheehan, 2015).

Applications of geographically distinct naked-eye observations of lunar eclipses to longitude determination became more successful in medieval Europe (Wright, 1922: 244–246; Woodward, 1987). Walcher of Malvern, the theological scholar and astronomer, obtained reliable results from simultaneous observations in England and Italy of the lunar eclipses of 30 October 1091, 18 October 1092 and those occurring in 1107–1112 (Wright, 1922: 244–246; Cortesão, 1969: 182–183). Later in the twelfth century, the astronomer and alchemist Roger of Hereford used simultaneous observations of the lunar eclipse of 12 September 1178 from Hereford, Marseilles and Toledo to calculate their longitudes with respect to that of Arin, the mythical centre of the Islamic world.

Columbus would thus have been familiar with the use of lunar eclipses as a means of longitude determination. In fact, he is known to have owned a copy of the 1474 *Kalendarium*, the astronomical almanac prepared by Johannes Müller von Königsberg (better known as Regiomontanus; 1436–1476), which predicted the times of lunar eclipses for the next 30 years. Columbus carried a copy of the almanac with him on at least his first transatlantic voyage (Morison, 1942: vol. I, 653–654; Morison, 1955). In fact, he reportedly observed a lunar eclipse while in the Americas:

> He declared also from the observation of his people that when in the year of our Lord 1494 there appeared an eclipse in the month of September, it was seen in Española four hours before that it was visible in Spain. (Thacher, 1903: 192, 195)

> On September 15th by the mercy of God they sighted an island which lies off the eastern end of Española … in the middle of a great storm he anchored behind this island … That night he observed an eclipse of the moon and was able to determine a difference in time of about five hours and twenty-three minutes between that place and Cádiz. (Colon and Keen, 1992: 48)

Columbus himself also reported to have observed this particular eclipse (West and Kling, 1991: 226–227):

> In the year 1494, when I was at the island of Saona, which is at the eastern end of the island of Hispaniola, there was an eclipse of the moon on the 14th of September, and it was found that there was a difference from there to the Cape of St Vincent in Portugal of five hours and more than one half, a claim he repeated in a letter of 7 July 1503 to King Ferdinand II of Aragon and Queen Isabella I of Castile:

> In the year ninety-four I navigated in twenty-four degrees [of latitude] to the westward to the end of nine hours, and I cannot be in error because there was an eclipse. (Jane, 1988: 82)

This was confirmed by Bartolomé de las Casas (c. 1484–1566), Bishop of Chiapas, in his *Historia de las Indias*:

> From the end of Cuba (that is seen in Hispaniola), which was called the End of the East, and by another name Alpha and Omega, he sailed westward from the southern part, until he passed the end of ten hours on the sphere, in such a way that when the Sun set to him, it was two hours before rising to those that lived in Cádiz, in Spain; and he says that there couldn't be

> any error, because there was an eclipse of the Moon on the 14th of September, and he was well prepared with instruments and the sky was very clear that night, (de las Casas, 1951: 309) and once again in a later chapter, although there de las Casas (1951: 395–396) reports a longitude difference of 5 hours 23 minutes West of Cádiz.

These historical records leave us with an unsatisfactory accuracy of Columbus' longitude determination. Pickering (1997) tried to disentangle the underlying causes of these discrepancies. The correct, modern longitude difference between Saona Island and Cádiz is 4 hours 10 minutes. If Columbus had based his measurements on Regiomontanus' (1489) almanac readings, he would have made a small mistake, since Regiomontanus' predicted time was 24 minutes late (see also Pickering, 1997)—he would thus have concluded that, at his observation point at Saona, he was 4 hours 34 minutes West of Cádiz.

The six longitude differences reported in the contemporary, primary records include five different measurements which are discrepant with respect to the actual difference by anywhere from 34 minutes to more than five hours. Local time could be measured to an accuracy of 10–15 minutes, while the eclipse onset might be uncertain by up to 5 minutes; these uncertainties do no add up to even the least discrepant measure of the local longitude.

Historians have long been suspicious of Columbus' reports (Morison, 1942; vol. II, 147). One cause for concern is a report that Columbus diverted to Saona to shelter from an approaching storm. This report was confirmed by his son Fernando, who added that the storm had already reached them by the time they reached their anchorage. But then, how could they have made accurate observations of the lunar eclipse timings, observations that are apparently lost today? And why did Columbus later suggest that the night was clear? Pickering (1997) has suggested that Columbus' motivation to report clearly fraudulent results was driven by ambition. In his *Libro de los Privilegios* (*Book of Privileges*, 1502) we read that Columbus asked the Spanish crown to confirm his dual status of 'Admiral of the Ocean Sea' and Viceroy of the New World. The Spanish sovereigns' confirmation specifically restricted his Admiral's privilege to only "the Ocean Sea in the region of the Indies" (Nader and Formisano, 1996: 73–74, 87, 151, 153), which may have motivated Columbus to represent the newly discovered lands as the Asian continent (Morison, 1942: vol. II, 140–141).

After all, as Columbus reminded his patrons in his letter of July 1503, the African/Eurasian landmass was thought to span from 12 to 15 hours in longitude, depending on whether one adopted Ptolemy's or Marinus of Tyre's calculations (Jane, 1988: 84). Therefore, Columbus likely anticipated that the remaining section of the globe, 9–12 hours, was covered by the extent of the 'Ocean Sea' between Europe and Asia, sailing westwards. Although this reasoning may explain why he claimed a longitude difference with respect to Cádiz of nine hours or more, the smaller differences reported by the other primary sources are still too discrepant from the actual value to be attributed to timing uncertainties alone.

These differences are likely owing to the misfortune that bad weather intervened in Columbus' attempt to observe the lunar eclipse of September 1494. In his *Historia de las Indias*, de las Casas (1951: 497) points out that Columbus measured the distance between the Canary Island of La Gomera and Dominica at 850 leagues. Adopting Columbus' conversion of 56⅔ leagues to a degree of longitude, this translates into four hours of longitude (although this conversion formally only applies at the Equator). Columbus himself did not report this small longitude, but his biographer Antonio Gallo did, perhaps based on a personal conversation with Columbus, his son Fernando or any of the sailors on board of the Admiral's ship. Note that the distance of 1142 leagues quoted in Columbus' *Diario*, when converted to degrees of longitude using the same conversion, translates to 5 hours 23 minutes—as reported by Fernando in his biography and by de las Casas in his *Historia*.

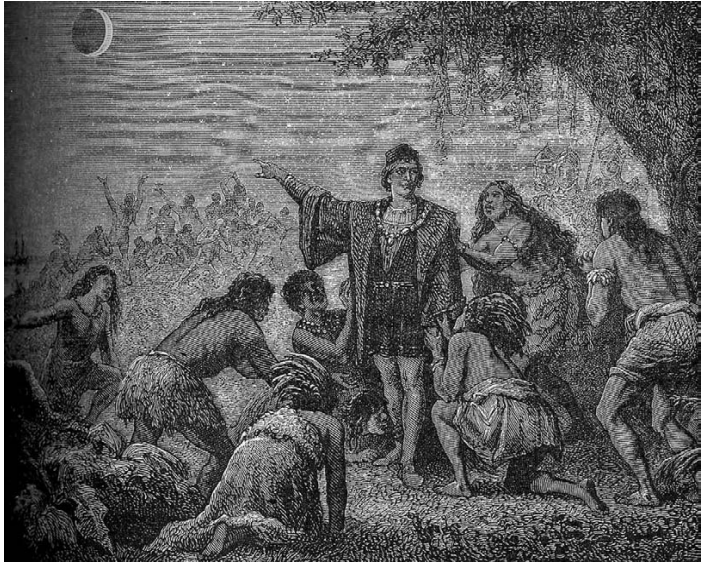

Pickering (1997) suggests that Columbus may have been unsatisfied with the 'short' transatlantic distance he had recorded in his *Diario*. At a later stage, he proceeded to add the East–West distance he sailed within the Caribbean islands to the earlier total of 1142 leagues, which would have corresponded to at least 31 leagues from the Bahamas to Cuba (Pickering, 1997), resulting in a total longitude difference with respect to Cádiz of 5 hours 31 minutes—similar to the difference he quoted in his *Book of Prophecies* (1501–1505).

**Figure 2**: Columbus on Jamaica during the lunar eclipse of 29 February 1504. (Source: Flammarion, 1879: 231, Fig. 86; *public domain*)

The numbers quoted in the *Book of Prophecies* are not all internally consistent, however: the longitude difference Columbus reported for the second lunar eclipse he observed from Jamaica in February 1504 (see Figure 2) is 7 hours 15 minutes West of Cádiz. The correct, modern longitude difference between both locations is 4 hours 44 minutes. In view of the fraudulent longitudes reported in relation to the September 1494 eclipse, it may well be that Columbus proceeded along similar lines here. In his July 1503 letter to the Spanish crown, he stated,

> When I set out thence to come to Española, the pilots believed that we were going to reach the island of San Juan [Puerto Rico], and it was the land of Mango [western Cuba], four hundred leagues more to the west than they said. (Jane, 1988: 98)

The distance from San Juan Island to Mango is the longest East–West distance Columbus recorded within the Caribbean archipelago. Added to the longest transatlantic distance he recorded on his first voyage, 1142¼ leagues (re-determined by Pickering, 1997), this yields a distance of 1542¼ leagues from Cádiz, corresponding to a difference of 7 hours 15 minutes in longitude—precisely the difference claimed by Columbus based on his purported observations of the 1504 eclipse (for a different explanation, involving Columbus having made a series of mistakes, see Olson, 1992).

## 2 FROM LUNAR ECLIPSES TO LUNAR DISTANCES

Lunar eclipses require close alignment of the Sun–Earth–Moon system. They occur only occasionally and so cannot be relied upon for routine longitude determination at sea. Instead, one could, in principle, use the projected position of the Moon with respect to a set of reference stars—as we already saw in the context of early Chinese longitude determinations.

This is the essence of the lunar distance method: given a sufficiently accurate representation of the Moon's path across the sky, one can calculate such lunar distances, or 'lunars', for any location on Earth and for specific times. At any other position, one can then determine the local time(s) at which one or more specific lunars occur. The difference between the tabulated time, in the sixteenth to nineteenth centuries most often that at Greenwich Observatory, and the local time at the location of the observer—determined from the altitude of the Sun, most often the Sun's local meridian passage—can be converted directly into a longitude difference. Although the principle of the lunar distance method is straightforward, we will see shortly that its practical implementation is anything but (for the method's mathematical basis and practical implementations, see e.g. Inman, 1819; Merrifield, 1884; van der Werf, 1977; de Man, 2017).

It is generally thought that the lunar distance method's first European publication can be traced back to the German mathematician Johann(es) Werner's *In hoc opere haec continentur Nova translatio primi libri geographiae Cl. Ptolomaei* (Nuremberg, 1514), which included a translation of Ptolemy's *Geographia*. However, Werner may not have been alone in pursuing longitude determination in the late Middle Ages. Around the time of his seminal publication, Ferdinand Magellan was preparing for his round-the-world voyage of 1519–1522. The explorer enlisted the services of the Spanish astronomer Andrés de San Martin (García and del Carmen, 1997) as chief pilot of his fleet, the *Armada del Maluco*. The astronomer brought with him Rui Faleiro's set of instructions related to longitude determination collected in Faleiro's *Regimento* (Lopes de Castanheda, 1554: 161), which included a version of the lunar distance method as one of three possible approaches to longitude determination (Randles, 1984: 9).

Randles (1984) has suggested that Faleiro may have drawn on a passage in the *Theori[c]a Planetarum*, most likely attributed (Pedersen, 1981) to the thirteenth-century Italian translator Gerard Cremonensis (also known as Gerard de Sabloneta),

> When the Moon is on the meridian, if you compare her position with that given in the lunar tables for some other locality, you may determine the difference in longitude between the place where you are and that for which the lunar tables were constructed by noting the differences in the position of the moon as actually observed and as recorded in the tables. It will not be necessary for you to wait for an eclipse. (Wright, 1922, transl.: 83–84)

Using Faleiro's proposed methods, de San Martin managed to accurately determine the longitudes of both Puerto San Julián in Patagonia (southern Argentina) and Homonhon Island in the eastern Philippines. It is likely that in doing so he relied predominantly on the lunar distance method. At Puerto San Julián, he used the Moon's conjunctions with the planets to derive a position 61° West of Seville. The modern longitude difference between both localities is 61.75°. This remarkably accurate longitude determination far exceeded that of contemporary navigators (Joyner, 1992). Upon their arrival at Homonhon Island on 16 March 1521, de San Martin determined a longitude difference of 189° "from the meridian", that is, from the line of demarcation, 47° West of Greenwich. By modern standards, Homonhon Island is located at 125° 43′ 47.3″ East. This implies that this 1521 longitude determination was once again remarkably accurate, to within 2° of the current value.

We will now return to Werner's seminal work. He appears to have been inspired by a letter from Amerigo Vespucci, the Italian explorer and Columbus' cartographer, to his patron, Lorenzo di Pierfrancesco de'Medici, referred to as 'letter IV' by Vespucci scholars. In that letter, the Italian claimed to have determined his longitude based on observations of celestial objects obtained on 23 August 1499 while on his second voyage in the Amazon river delta.

Indeed, in a sixteenth-century copy of a fragmentary letter known as the *Ridolfi Fragment*, discovered in 1937 by Professor Marques Ridolfi in the Conti archives (Conti de Ottomano Freducci, Archivio di Stato, Firenze, Italy) and possibly dating from 1502, Vespucci wrote to de'Medici,

> Briefly to support what I assert, and to defend myself from the talk of the malicious, I maintain that I learnt this [my longitude] by the eclipses and conjunctions of the Moon with the planets; and I have lost many nights of sleep in reconciling my calculations with the precepts of those sages who have devised the manuals and written of the movements, conjunctions, aspects, and eclipses of the two luminaries and of the wandering stars, such as the wise King Don Alfonso in his *Tables*, Johannes Regiomontanus in his *Almanac*, and Blanchinus, and the Jewish rabbi Zacuto in his *Almanac*, which is perpetual; and these were composed in different meridians: King Don Alfonso's book in the meridian of Toledo, and Johannes Regiomontanus' in that of Ferrara, and the other two in that of Salamanca. It is certain that I found myself, in a region that is not uninhabited but highly populated, 150 degrees west of the meridian of Alexandria, which is eight equinoctial hours. If some envious or malicious person does not believe this, let him come to me, that I may affirm this with calculations, authorities, and witnesses. And let that suffice with respect to longitude; for, if I were not so busy, I would send you full detail of all the many conjunctions which I observed, but I do not wish to become tangled in this matter, which strikes me to be the doubt of a literary man, and not one which you have raised. Let that suffice. (Vespucci, 1992: 38–39)

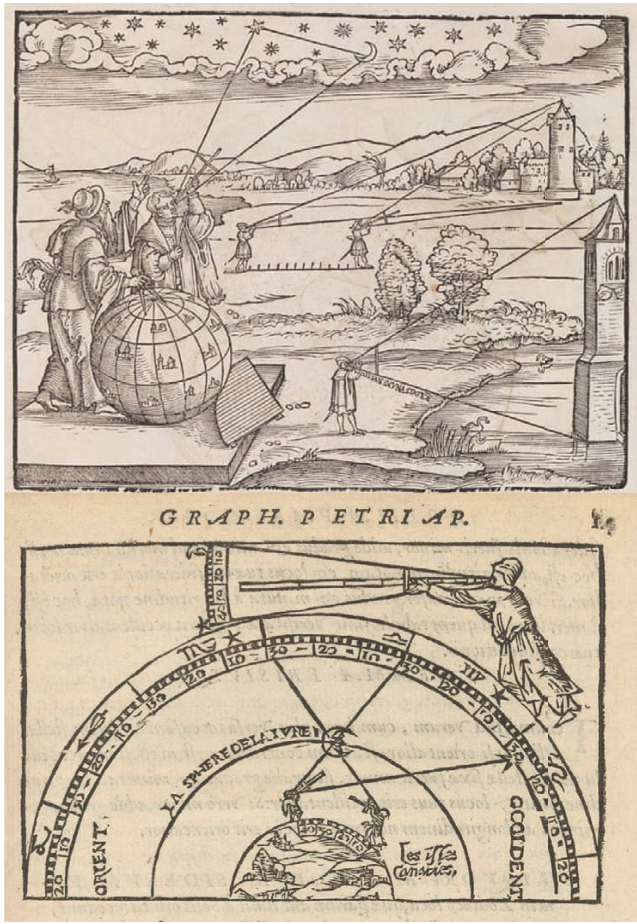

**Figure 3**: Illustrations from (*top*) the title page of Apianus' *Introductio Geographica* (Ingolstadt, 1533; *public domain*), and (*bottom*) a folio from the *Cosmographia Petri Apiani* (1524; *public domain*), both showing the lunar distance method in practice. Note the cross staffs employed by all observers.

Werner had carefully studied Regiomontanus' astronomical treatises. His subsequent observations of the orbital motion of a comet from 1 to 24 June 1500 laid the basis for what we now know as the lunar distance method. Although an ordained priest (vicar) at Nuremberg since 1497, "his pastoral duties were rather limited, [so] he devoted much time to scientific study" (Folkerts, 2020). Following Regiomontanus' teachings, his 1514 manuscript discusses longitude determination on the basis of both lunar eclipses and the Moon's position relative to the fixed stars or the Sun:

> Our aim is to find the distance in longitude between two distant places. The geographer will be in one of these places and will measure with a cross-staff [as already used for latitude determination at the time] the distance of the Moon from a star on the Ecliptic [Earth's equatorial plane with respect to its orbital plane]. If then we divide this distance by the velocity of the Moon per hour, we will know at what time in the future the Moon will be in conjunction with this body. (Werner, 1514)

Unfortunately, however, during Werner's lifetime the celestial ephemerides were known to insufficient precision to allow this method to work. Werner's method was discussed at length by Petrus Apianus, the German humanist and mathematician. Apianus is perhaps best known for his 'volvelles' or wheel charts, also known as 'Apian wheels', an early form of analogue computers composed of rotating paper parts. He was clearly enamoured by Werner's lunar distance method (e.g., Gislén, 2018, 2019). He even illustrated the method on the title page of his *Introductio Geographica* (Ingolstadt, 1533): see Figure 3. Apianus' first major work, *Cosmographia seu descriptio totius orbis* (Landshut, 1524), was the pre-eminent sixteenth-century textbook for basic astronomy, surveying and cartography. Subsequent editions were improved and expanded by Jemme Reinierszoon, the Professor of Medicine at the University of Leuven better known as Gemma Frisius. The latter added numerous appendices to the *Cosmographia*, including one outlining the use of clocks for determining one's longitude (Pogo, 1935).

Apianus' lunar distance method employed the cross staff (see Figure 4). However, although useful for measuring latitudes, at the time the accuracy of the cross staff was insufficient for use with the lunar distance method. One needed better lunar almanac tables and the development of the octant. This latter development took until the early 1730s to mature, with the (reflecting) octant's first practical implementation by the English mathematician John Hadley (aided by his brother George) and Thomas Godfrey, an inventor based in the American colonies. The octant became the instrument of choice for navigation at sea by the 1750s:

> The Instrument is design'd to be of Use, where the Motion of the Objects, or any Circumstance occasioning an Unsteadiness in the common Instruments, renders the Observations difficult or uncertain. (Hadley, 1731: 147; see also Figure 5)

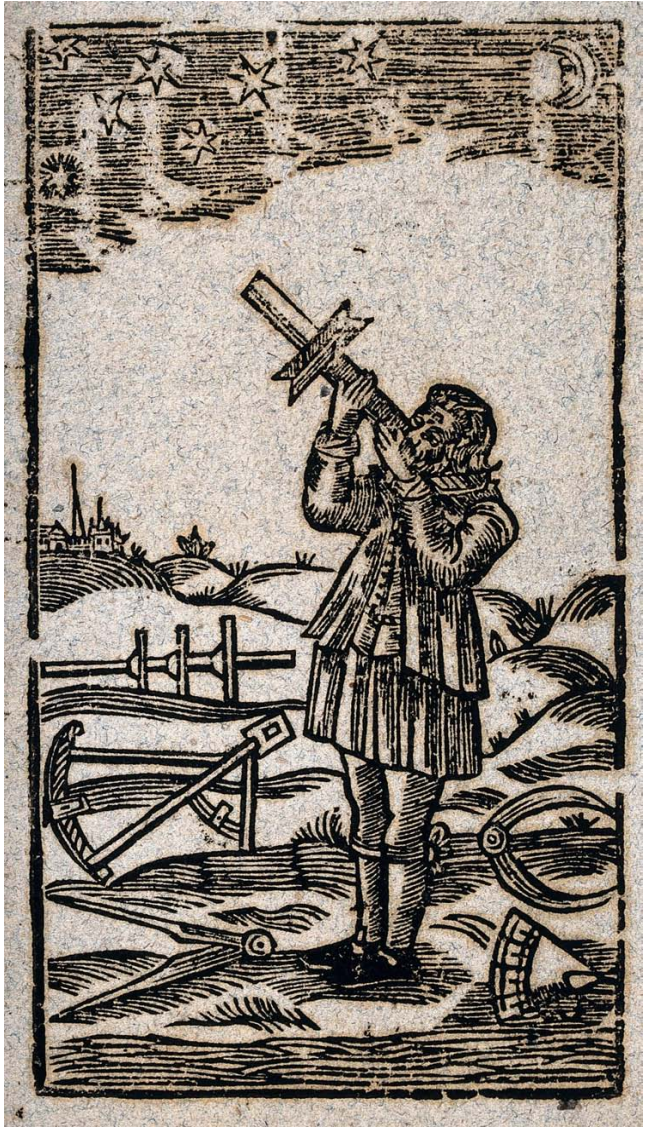

**Figure 4**: Application of the cross staff (Jacob's staff) for angular measurements between celestial objects (woodcut, seventeenth century). Other geometric measurement instruments are shown on the ground in larger-than-natural size. On the left, a different type of cross staff is shown. (Credit: Wellcome Collection. Attribution 4.0 International; CC BY 4.0)

Independently from Werner and Apianus, in 1634 the French mathematician and astronomer Jean-Baptiste Morin claimed that he had discovered a new way of finding longitude, commenting that, "I do not know if the Devil will succeed in making a longitude timekeeper but it is folly for man to try" (Marguet, 1917: 7). Cardinal Richelieu, King Louis XIII's chief minister, established a commission to examine this claim, the *Conférence des longitudes* (Bigourdan, 1916; Pares, 1977). The body was composed of Henri-Auguste de Loménie, Count of Brienne, other government officials, naval officers and mathematicians, including Blaise Pascal's father Étienne, later treasurer of France, Pierre Hérigone, Claude Mydorge, Jean de Beaugrand and Jean Boulenger. They concluded that Morin's approach, which was based on measuring absolute time by determining the position of the Moon relative to the stars and which took into account atmospheric refraction as well as the lunar parallax (small positional shifts depending on the observer's location), was simply a more sophisticated version of the common lunar distance method (Bigourdan, 1916).

Meanwhile, in the early seventeenth century proposals for the development of viable methods of longitude determination at sea were actively considered by the Spanish *Conseja de Indias* (e.g., de Grijs, 2020). In 1567, King Philip II announced a significant monetary reward for anyone who could solve the intractable longitude problem. Upon his accession to the Spanish throne in 1598, King Philip III followed his father's lead by sponsoring a significantly more generous longitude prize.

Hoping for a share of this Spanish longitude prize, the Flemish cartographer and astronomer Michael Florent van Langren, seconded as cosmographer and mathematician to the Spanish court, pursued a method of longitude determination based on the appearance and disappearance of lunar features, particularly of the rising and setting of peaks and craters. He realised that as the lunar phases progress from new to full moon, the Sun progressively illuminates different lunar features from East to West, and *vice versa* from full to new moon. In 1625 he presented his method as a possible solution to the longitude problem to princess Isabella Clara Eugenia, daughter of Phillip II (Navarro Brotons, 2018; de Grijs, 2020). Despite initial royal encouragement (de Grijs, 2020), van Langren's approach ultimately failed, most importantly because lunar features appear gradually rather than instantaneously.

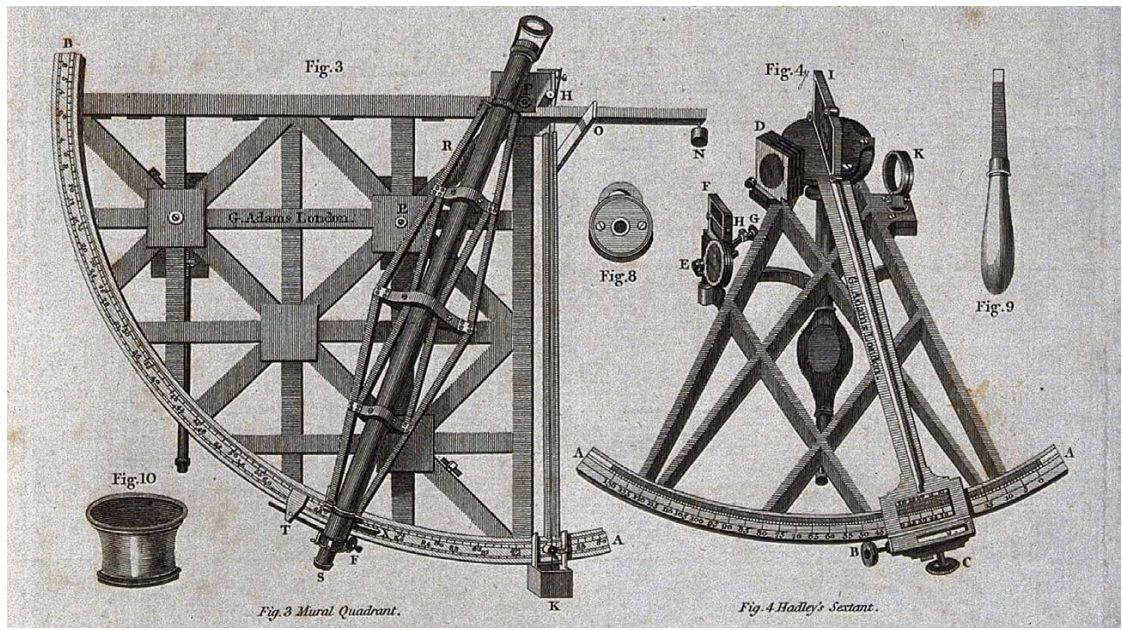

**Figure 5**: (*left*) Mural quadrant ('Fig. 3'); (*right*) Hadley's sextant ('Fig. 4'). Engraving (detail). (Credit: Wellcome Collection. Attribution 4.0 International; CC BY 4.0)

Around the same time, the Spanish philosopher and mathematician Juan Caramuel y Lobkowitz presented an altogether novel and systematic approach to longitude determination based on lunar positions. In 1625, had Caramuel joined the Cistercian Order at the Monastery of La Espina (García Camarero, 2018) under the patronage of Pedro de Ureña. Although he was blind, de Ureña had independently devised his own version of the lunar distance method, but he passed away before he managed to get his method published (Sanhuesa Fonseca, 1999; de Barcelos e Coles, 2014). He left the printing and dissemination of his invention to Caramuel (Velarde Lombraña, 1982: 505; for an in-depth discussion, see de Grijs, 2020).

Tabulated lunar distances reflect the Moon's position as it would be seen by a geocentric observer, that is, an observer located at the centre of the Earth. One must therefore correct for the observer's location on the Earth's surface, which may require an angular shift of up to a degree. In practice, a shipboard observer would measure the projected angular distance between the Moon's sharp limb and his[3] selected reference object using a sextant or an octant. He would then need to correct the observed separation as well as the Moon's altitude for the projected size of the Moon on the date of the observation (Dunlop and Schufeldt, 1972: 409), a quantity tabulated in contemporary almanacs. The combined correction for atmospheric refraction and lunar parallax—usually also provided in tabular form—is a relatively straightforward trigonometric function involving the observed lunar distance and the respective altitudes of the Moon and the reference star(s) (Van Flandern and Pulkkinen, 1979; see also Schlyter, n.d.), for which navigators would usually use pre-calculated mathematical tables. This process is known as 'clearing' lunar distances. The objects' altitudes should be measured both before and after measuring the lunar distance itself or, if possible, simultaneously (Stark, 2010; x–xiii). Finally, the observer would consult his almanac again to read off the exact time at Greenwich Observatory implied by the measured lunar distance. Combined with his local time, independently obtained from observations of the Sun or bright references stars crossing the local meridian, this would then result in an operational longitude determination.

Despite this apparently straightforward approach, Morin and the *Conférence des longitudes* were in dispute about the new method's feasibility until 1639, given that ephemeris tables of the requisite accuracy were as yet non-existent. During this period, Morin proposed to establish an observatory in Paris (Delambre, 1821: 242; see Figure 6) to obtain regular observations of the positions of the Moon and selected reference stars for the dual purposes of longitude determination and cartographic corrections (Wolf, 1902: 2). Paris Observatory was eventually established in 1667, although Morin's influence in getting the project off the ground is unclear (Deias, 2017). Meanwhile, Richelieu died in 1642. In 1645, his successor, Cardinal Jules Raymond Mazarin, gave Morin 2000 *livres* for his efforts, but his proposed method was still deemed impractical.

In 1674, a French explorer calling himself Le Sieur de Saint-Pierre, most likely Jean-Paul Le Gardeur (de Grijs, 2017: Ch. 6), claimed to have solved the longitude problem based on application of the lunar distance method (Forbes, 1976). Meanwhile, in England, King Charles II had been receiving ever increasing numbers of proposals from scholars and opportunists alike claiming to have solved the longitude problem. This development prompted him to appoint a Longitude Commission to validate these proposals,

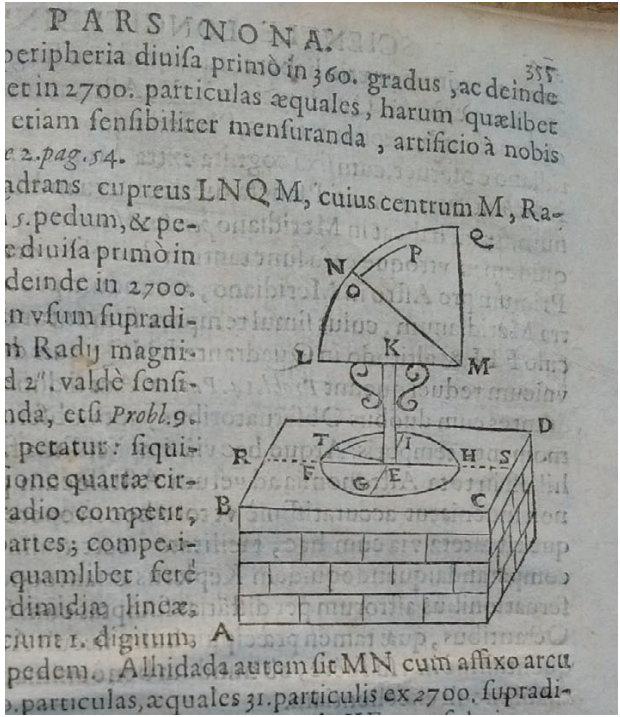

**Figure 6**: Detail of the fixed observation structure proposed by Jean-Baptiste Morin (Morin, 1634, Ch. IX, p. 355; Bibl. Nat., V-18023; *public domain*).

… that he [de Saint-Pierre] hath found out the true knowledge of the Longitude, and desires to be put on Tryall thereof; Wee having taken the Same into Our consideration, and being willing to give all fitting encouragement to an Undertaking soe beneficiall to the Publick … hereby doe constitute and appoint you [the Commissioners], or any four of you, to meet together … And You are to call to your assistance such Persons, as You shall think fit: And Our pleasure is that when you have had sufficient Tryalls of his Skill in this matter of finding out the true Longitude from such observations, as You shall have made and given him, that you make Report thereof together with your opinions there-upon, how farre it may be Practicable and usefull to the Publick.[4]

On 4 March 1675, the King signed a royal warrant to appoint John Flamsteed as "our astronomical observator" and assistant to the Royal Commission. He proceeded to undertake an observational programme to test the viability of de Saint-Pierre's proposal. Although he acknowledged the method's potential, Flamsteed also highlighted its practical infeasibility given the lack of detailed tables of the positions of the fixed stars relative to both the Sun's annual path and the Moon's orbital parameters.

Meanwhile, while compiling the first edition of his *Philosophiæ Naturalis Principia Mathematica* (London, 1687), Isaac Newton (1687: 434) hoped to use the Moon's path across the sky for longitude determination at sea in an attempt to combine the best aspects of astronomy and accurate timekeeping. Newton understood that the success of the lunar distance method depended on an accurate understanding of the lunar motions. These are fairly complex, however, given that the Moon's orbit is determined by three-body gravitational interactions among the Sun, the Earth and the Moon. Large numbers of careful and precise observations of the Moon's motion relative to the fixed stars are required to accurately trace its movement across the sky. The global pattern, driven by precession of the lunar orbit relative to the ecliptic plane, repeats only once every 18.6 years while also exhibiting repeating minor perturbations and fluctuations, together posing a major problem for Newton and his contemporaries (Smith, 1999).

As we will shortly see, despite the significant efforts he undertook in preparation for the three editions of his *Principia*, Newton did not manage to obtain a satisfactory solution to the problem. He continued to refine his treatment of the lunar motion and its application to longitude determination until the publication of the *Principia*'s second edition (Newton, 1713; Cook, 2000). Although Newton had apparently already invented an innovative reflecting octant around 1699, which was meant to measure distances between the Moon and stars while at sea, and hence to help him in his modeling efforts, a description of the instrument (Newton, 1699) was only found among the papers of Edmond Halley, Britain's second Astronomer Royal, long after Newton's (and Halley's) death:

> By this instrument the distance of the moon from any fixed star is thus observed; view the star through the perspicil [telescope eyepiece] by the direct light, and the moon by the reflexed, (or on the contrary;) and turn the index till the star touch the limb of the moon, and the index shall show on the brass limb of the instrument the distance of the star from the moon's limb; and though the instrument shake by the motion of the ship at sea,[5] yet the moon and star will move together as if they did really touch one another in the heavens; so that an observation may be made as exactly at sea as at land. And by the same instrument, may be observed exactly the altitudes of the moon and stars, by bringing them to the horizon; and thereby the latitude and times of observation may be determined more exactly than by the ways now in use. (Newton, 1742)

## 3 FLAMSTEED'S MURAL ARC

Flamsteed realised that he would need years of observational data, obtained with large instruments fitted with telescopic sights, to make the lunar distance method a practical reality. On 11 September 1689, he inaugurated a new mural arc at Greenwich Observatory, allowing for more precise observations. His first task following verification of the arc's position and calibration of its performance was to determine the equinox, that is, the date on which day and night are approximately equally long anywhere on Earth, the Observatory's latitude, the obliquity (inclination) of the Earth's ecliptic plane and other fundamental measurements required to correctly determine the positions of the fixed stars.

Flamsteed also determined true solar, lunar and planetary motions. Newton was keen to use these as the basis of his lunar distance method, but it took a number of years before Flamsteed was prepared to make his results public. Flamsteed's mural arc, used alongside an accurate pendulum clock, allowed him to measure zenith distances of objects crossing the local meridian, as well as the exact sidereal times when this happened. In turn, these measurements enabled him to calculate the objects' right ascensions and declinations, that is, their accurate positions on the sky.

It is clear that Newton was struggling to make sense of the observations he had access to. Ultimately, Newton's failure to solve the complex Sun–Earth–Moon three-body problem led to a significantly deteriorated relationship with Flamsteed, whom Newton accused of ineptitude. Nevertheless, one should keep in mind that Newton was the first to attempt to solve the three-body system's equations of motion, and that his results were nevertheless remarkably successful. The three-body problem was eventually solved numerically by the Swiss polymath Leonhard Euler,[6] who received a reward of £300 from the British Board of Longitude (see below) in May 1765 "for Theorems furnished by him to assist Professor [Tobias] Mayer in the Construction of Lunar tables". (Howse, 1998: 408). I will shortly discuss Mayer's contributions (see Section 4).

The full body of Flamsteed's observations was eventually published posthumously in the second volume of his *Historia Coelestis Britannica* (1725), edited by his wife Margaret. During his lifetime, he had put off publishing his observations until completed, despite being pressured to do so by, among others, Newton and Halley. Around 1683, Halley actually contributed observations of occasional lunar occultations (eclipses) of some bright stars, as well as their 'appulses' (close encounters on the sky). He had hoped to use these measurements as accurate timing devices (Halley, 1731), that is,

> … he had found it needed only a little practice to be able to manage a five or six foot telescope capable of shewing the appulses or occultations of the fixed stars by the moon on ship-board in moderate weather, especially in the first and last quarter of the moon's age, when her weaker light does not so much efface that of the stars … Now the motion of the moon being so swift as to afford us scarce ever less than two minutes for each degree of longitude, and sometimes two and a half, it is evident that we were able perfectly to predict the true time of the appulse or occultation of a fixed star in any known meridian, we might, by comparing therewith the time observed onboard a ship at sea, conclude safely how much the ship is to the eastward or westward of the meridian of our calculus. … [but] the best tables then extant [in 1715] were too imperfect for this purpose; but that the errors of the tables returning to pretty near the same quantity after a period of 18 years and 11 days, or 223 lunations, the tables might be corrected at any time from observations made at that distance of time in an antecedent period, provided such were made. (Mayer, 1771)

Unfortunately, the occultation method turned out to be too cumbersome for practical application given the scarcity of bright stars and their infrequent lunar occultations.[7] It was deemed impractical to train maritime navigators to recognise a useful set of dimmer stars.

The three volumes of Flamsteed's *Historia Coelestis* contained more than 28,000 observations made with the mural arc between 11 September 1689 and 27 December 1719, which provided 2935 stellar positions more accurately than published in any previous compilation. Although the catalogue was not entirely free of errors, the tabulated stellar, lunar and planetary positions were sufficiently accurate to make the lunar distance method a viable navigation approach.[8] However, one still needed to correct the motions of the Sun and the Moon and somehow measure the Moon's position with respect to the fixed stars while at sea.

Meanwhile, in the first edition of his *Principia* (Book I, Proposition LXVI; Book III), Newton had discussed the Moon's motion in general terms, concluding that understanding the effects of the Sun's gravity was paramount. He expressed the Sun's gravitational pull as a perturbation in the Moon's Earth-centred acceleration. His treatise from 1702 aided him in improving his description of the Moon's motion, published in the *Principia*'s second edition (1712). The description in the third edition was largely consistent with that published in the previous edition, although Newton included an explanatory addendum by Machin (1726) detailing the motion of the Moon's orbital nodes (Cook, 2000). At the same time, Flamsteed had used Newton's insights from 1702 to prepare a set of tables that included some lunar observations; however, these were not published until 1746, posthumously, and only in France (Le Monnier, 1746: 251).

Separately, Halley had played an important role in firming up the observational basis of Newton's lunar orbital theory. By 1719, he had prepared—but did not publish—a large number of observational tables pertaining to a range of celestial bodies, including the Moon, representing Newton's version of his theory as published in the *Principia*'s second edition. Between 1721 and 1739, Halley obtained numerous detailed observations of the Moon. Again, however, he did not publish his observational data, although halfway through his 18-year observation cycle, in 1731, he offered some early insights into the accuracy of Newton's theory (Halley, 1731). Halley suggested that the Moon's positions as predicted by Newton's theory agreed with his own observations to within 2 arcmin (standard deviation, with outliers of up to 8 arcmin) of projected separation (Cook, 2000), and that his own observations agreed well with those of Flamsteed and others, which had been obtained several lunar cycles previously. Halley's full tables were eventually published posthumously (Halley, 1749, 1752).

Meanwhile, on 9 July 1714 Queen Anne had given her royal assent to *An Act for providing a Public Reward for such Person or Persons as shall discover the Longitude at Sea*. As a result, 22 'Commissioners for the Discovery of the Longitude at Sea', popularly known as the 'Board of Longitude', were initially appointed to adjudicate claims. The British Longitude Act included a reward of up to £2000 for work that appeared promising and £10,000 for a demonstrably practical method that was successful in determining one's longitude at sea with an accuracy down to one degree of a great circle, equivalent to 60 nautical miles at the Equator. The monetary reward would be increased to £15,000 for accuracies of 40 arcmin and to £20,000 for methods with an accuracy of half a degree or better.

The Longitude Act did not include a role for an actual Board of Longitude nor any formal procedure for submitting proposals or demonstrations. All interested parties were expected to contact the Commissioners or the Admiralty through published pamphlets, broadsheets or newspapers, by sending letters or in person. (The *de facto* Board of Longitude did not meet in person until 1737; Baker, 2015.) The Queen's endorsement of the 1714 Longitude Act led to a flurry of activity, generating numerous proposals from genuine scientist-scholars, lunatics and those hoping to become wealthy easily.

The increasing importance of Flamsteed's observations, combined with the frenzied activity triggered by the passing of the Longitude Act, led Robert Wright to prepare his *Viaticum Nautarum* (*Sailor's Vade Mecum*; Wright, 1726, 1732). Wright (1726) emphasised the need for a practical quadrant and included images of such an instrument, a full nine years prior to the development of Hadley's reflecting octant. His follow-up submission of 1732 included lunar tables required to "make a complete System of all that is yet wanting in Navigation". Yet, despite his careful preparation, the Commissioners of the Board of Longitude seem to have ignored his submission:

> I hope I may be permitted to say without boasting, that I cannot think this so petty, trivial, jejeune, and insignificant a Performance, that it ought to be thrown a-side unregarded, and buried in Silence. (Barrett, 2013)

**4 LARGE-SCALE PRACTICAL APPLICATIONS**

It took until the 1750s before sufficiently accurate lunar tables became available for large-scale practical use. While employed at the Göttingen University Observatory, the German astronomer Tobias Mayer was instrumental in developing improved lunar and solar ephemeris tables (Forbes, 1971). Mayer was once referred to by Euler as "undoubtedly the greatest astronomer in Europe", while the French astronomer Jean Delambre heaped even more lavish praise on the cartographer, referring to him as "universally considered … one of the greatest astronomers not only of the eighteenth century, but of all times and of all countries" (Delambre, 1827).

In 1752, Mayer designed a novel 'repeating circle', allowing him to obtain more accurate measurements, as we learn from the Spanish astronomer José de Mendoza y Ríos (1801):

> As the reflecting instruments employed at sea are supported by hand, their weight and scale are limited within a narrow compass; and it seemed very difficult to obviate, by any expedient, the inconveniences arising from the smallness of their size, while it was impossible to increase it. The celebrated Tobias Mayer contrived, however, a method to determine, at one reading, instead of the simple angle observed, a multiple of the same angle; and, by this means, the instrument became, in practice, capable of any degree of accuracy, as far as regards the above mentioned errors. His invention is essentially different from the mere repetition of the observations ...

From 1753, he employed his novel instrument to improve and further develop the lunar and solar ephemeris tables he needed to obtain better longitude determinations in support of his cartographic work. (After 1778, Mayer's repeating circle was perfected by the French astronomer Jean-Charles de Borda, whom we will encounter shortly.) However, Mayer was reluctant to submit his work to the British Board of Longitude since he did not believe the Board would award its Longitude Prize to a foreigner. Nevertheless, in 1757 he was eventually convinced to submit an amended set of lunar tables. His tables were sufficient to determine longitude at sea to better than half a degree, as noted by Nevil Maskelyne, the fifth British Astronomer Royal, in *The British Mariner's Guide* (1763):

> The Tables of the Moon had been brought by the late Professor Mayer of Göttingen to a sufficient exactness to determine the Longitude at Sea to within a Degree, as appeared by the Trials of several Persons who made use of them. The Difficulty and Length of the necessary Calculations seemed the only Obstacles to hinder them from becoming of general Use.

In 1760, while writing a preface to his tables, Mayer had commented,

> I am the more unwilling my tables should lie any longer concealed; especially as the most celebrated astronomers of almost every age have ardently wished for a perfect theory of the Moon ... on account of its singular use in navigation. I have constructed these tables ... with respect to the inequalities of motions, from that famous theory of the great Newton, which that eminent mathematician Eulerus first elegantly reduced to general analytic equations. (O'Connor and Robertson, 2008)

Yet it took until 1767 before his tables and the underlying theory (*Theoria lunae juxta systema Newtonianum*) were published, well after his death. Meanwhile, James Bradley, Britain's third Astronomer Royal, had carefully checked Mayer's tables. In his first letter to the Secretary of the Admiralty, of 10 February 1756, the Astronomer Royal wrote

> … that he had carefully examined Mr. Professor Mayer's theory and tables of the moon's motions, and other papers relating to the method of finding the longitude at sea, and compared several observations made (during the last five years) at the Royal Observatory at Greenwich, with the places of the moon computed by the said tables; and in more than 230 comparisons, which I have already made, I did not find any difference so great as 1′½ between the observed longitude of the moon and that which I computed by the tables: and although the greatest difference which occurred is, in fact, but a small quantity; yet as it ought to be considered as arising partly from the error of the tables, it seems probable, that during this interval of time, the tables generally gave the moon's place true within one minute of a degree. A more general comparison may, perhaps, discover larger errors; but those which I have hitherto met with being so small, that even the biggest could occasion an error of but little more than half a degree in longitude, it may be hoped, that the tables of the moon's motions are exact enough for the purpose of finding at sea the longitude of a ship, provided that the observations that are necessary to be made on ship-board can be taken with sufficient exactness. The method of finding the longitude of a ship at sea by the moon, hath been often proposed, but the defects of the lunar tables have hitherto rendered it so very imperfect and precarious, that few persons have attempted to put it in practice; but those defects being now in great measure removed, it may well deserve the attention of my Lords Commissioners of the Admiralty (as also of the Board of Longitude) to consider what other obstacles yet remain, and what trials and experiments may be proper to be made on ship-board, in order to enable them to judge whether observations for this purpose can be taken at sea with the desired accuracy. (Mayer, 1771; Rigaud, 1831: 84–85)

In a follow-up letter, dated 14 April 1760, he attested to the success of the method's application at sea:

> I computed the ship's longitude from each of the observations made by captain Campbell, and, upon comparing the results of several that were taken near the same time, and under the like circumstances, it appeared, that in general the observer was not liable to err more than one minute in judging of the apparent contact of the moon's limb and the object with which it was compared. Now this being nearly the same error that would be found to obtain if the like observations were to be made with the same instruments on land, it may hence be inferred, that in moderate weather the motion of the ship is not otherwise an impediment in this sort of observations, than as it renders the repetition of them more tedious and troublesome to the observer, which yet ought by no means to be omitted; because if each single observation be liable to an error of a minute only, by taking the mean of five or six, the error on this head may be so far diminished as to be of small moment. (Mayer, 1771; Rigaud, 1831: 86–88)

Shortly afterwards, Maskelyne used the tables on his voyages to St. Helena in 1761 and Barbados in 1763. The fact that he became a lifelong convert to the use of the lunar distance method following these voyages speaks for itself. In fact, in a memorandum to the Commissioners of the Board of Longitude of 9 February 1765, Maskelyne argued

> … that the longitude deduced from observations made by himself and others, with the help of Mr. Mayer's printed tables, always came within a degree; but as I am informed that Mayer's last manuscript tables are much more exact than the printed ones, it may be presumed that the longitude deduced from them will come considerably within a degree. (Mayer, 1771)

At Maskelyne's request, several experienced mariners attended the Board's next meeting, providing supporting evidence in the form of their journals and other documents,

> … that they had determined the longitude of their respective ships, from time to time, by observations of the moon, taken in the manner directed by the aforesaid book, and found the said observations easily and exactly to be made, and that the longitude resulting always agreed with the making of land (near the time of making the observations) to one degree; that they could make the observations in a few hours, not exceeding four hours, and are of opinion, that of a Nautical Ephemeris was published, this method might be easily and generally practised by seamen. (Mayer, 1771)

In response and as a silver lining to the delay in getting Mayer's tables published, the Board subsequently awarded his widow £3000 in return for his papers, although they had initially committed a larger sum, at least in principle:

> The Board, having taken the matter into consideration, came thereupon to the following Resolution, viz. Resolved, That it is the opinion of this Board, upon the Evidence given of the utility of the late Professor MAYER's Lunar Tables, that it is proper the said Tables should be printed; and that application should be made to Parliament for power to give a sum, not exceeding 5000 *l*., to the widow of the said Professor, as a reward for the said Tables, part of which have been communicated by her since her husband's decease; … (*Nautical Almanac*, 1834: iv) while the Board also approved the establishment of an annual 'Nautical Ephemeris' under the auspices of Maskelyne so as to allow mariners to pursue lunar distance determination at sea. (Around the same time, the Board also acknowledged Charles Mason's lifelong work to improve the existing lunar tables, particularly those produced by Mayer, but it took numerous petitions by his family to obtain posthumous remuneration of £750, in 1787.)

Although the £3000 offered to Mayer's widow was a considerable amount, equivalent to about £419,000 in 2019 pounds sterling,[9] it fell far short of the total amount available for the British Longitude Prize, that is, £20,000. In fact, in 1767 Parliament had awarded John Harrison half of the total prize money as reward for the successful sea trials of his marine timekeeper, H4. Bradley, who had also put significant effort into perfecting the lunar ephemeris tables, is said to have later told Harrison that he and Mayer would have shared the prize money but for Harrison's "blasted watch".

**5 THE *NAUTICAL ALMANAC***

The Board of Longitude's directive to establish an annual ephemeris publication led to a flurry of activity (Howse, 1989). Many entrepreneurial characters saw an opportunity to make some money from producing updated or novel lunar tables. They were not commissioned by government hydrographic offices or national navies, but their developers were, in essence, sole contractors. Some of their publications were subsidised but often still quite expensive given that the costs were almost directly proportional to the number of pages printed. In reality, many of the 'new' methods proposed were, in fact, existing methods in a new guise, often through cosmetic rearrangements of the equations involved or by computing a few more terms—mostly of negligible importance—of the lunar distance corrections required.

Maskelyne employed Israel Lyons the Younger, the mathematician, botanist and astronomer, to work on the *Nautical Almanac and Astronomical Ephemeris* and the *Tables Requisite to be used with the Astronomical and Nautical Ephemeris*. The Board of Longitude's minutes of 28 May 1765 confirm Lyons' appointment:

> Resolved
> That a Method Invented by Mr Israel Lyons of the Town of Cambridge & now delivered in by Mr [Anthony] Shepherd for finding the corrections of the Refraction of Parallax. And also a method, for the like purpose, invented by Mr George Witchell ... be sent, as soon as Mr Witchell shall deliver in his ... and that Mr Witchell & Mr Lyons be desired to deliver in Estimates of the Expence which will attend the making & publishing the Tables … (Board of Longitude, 1765: 8)

and on 13 June 1765 the Board agreed

> … that Mr Israel Lyons & Mr George Witchell (who were proposed as proper persons by the Professors), be employed by the Royal Astronomer to Calculate an Ephemeris for the year 1767 agreeable to the plan given in to the Last Board. (Board of Longitude, 1765: 101)

Lyons and Witchell used a 'short' or 'approximate' approach to calculating lunar distance corrections, in essence only including a limited number of terms in the correction equations. Such methods were most useful for shipboard practice. The first 'approximate' method employed was that of the French astronomer, Abbé Nicolas-Louis de Lacaille (de la Caille; e.g., Warner, 2002; Glass, 2012). His lunar distance tables, based on the orbital theory of the Newtonian French astronomer, mathematician and geophysicist Alexis Claude Clairaut (1752), were the first that were sufficiently accurate for time and longitude determinations at sea (de Lacaille, 1755b).

De Lacaille's development of his lunar distance tables and the adoption of his innovative graphical method by the French and, subsequently, the British scientific establishments had not all been plain sailing however. In 1742 de Lacaille was given the responsibility to calculate ephemeris tables for publication in the *Ephémérides des mouvements célestes*. Although he intended to include a description of the lunar distance method in the next volume of these ephemeris tables, covering 1745–1746, he abandoned that idea when he was advised that Pierre-Louis Moreau de Maupertuis, "responsible for the improvement of the navigation" and his superior at the French *Académie des sciences*, was about to publish his *Astronomie nautique* (1743). Contrary to de Lacaille's expectations, de Maupertuis did not discuss the lunar distance method, however (Boistel, 2015).

Subsequently, on a voyage in 1750–1754 to the Cape of Good Hope, de Lacaille collaborated with Jean-Baptiste d'Après de Mannevillette, a French naval officer associated with the *Compagnie des Indes*, who became the first to determine longitudes at sea using the lunar distance method, with an accuracy of 5 to 15 'lieues marines' (marine leagues: 25 to 45 km; d'Après de Mannevillette, 1754; Boistel, 2015). The voyage proved fruitful and productive, as we learn from de Lacaille's own report:

> During this sea voyage, I occupied myself in making trials of the method of observing longitude at sea by means of the distances of the Moon from some zodiacal fixed star. Following my departure from France, I made numerous investigations to facilitate the practice of the method proposed by Mister Halley.[10] I recognised that it was useless to look for another way of using the Moon for the longitude; that it was solely a question of making the calculation easy for ordinary sailors. (de Lacaille, 1755a)

Nevertheless, de Lacaille's proposals (which also included compiling a nautical almanac-type compilation of lunar distances to the Sun and other bright stars, an endeavour Maskelyne would pursue a decade hence) were blocked at home by jealousy, rivalry and competition (e.g., Boistel, 2015). Upon his return to France, in 1755 he was able to take ownership of his idea to supply pre-calculated tables of lunar distances, which in 1759 he supplemented by the introduction of a graphical method meant to avoid the need for lengthy and complex calculations at sea.

De Lacaille's graphical method was based on an earlier suggestion by the Jesuit priest Paul Hoste from 1692, worked out in more detail by a certain Mr. Griffon, a teacher of navigation (Boistel, 2015). One would draw a circle representing the celestial sphere with the pole and the Equator at right angles to each other, situating the observer in its centre. Local time resulted from plotting the Sun's path across the firmament, based on tabulated data in contemporary almanacs. De Lacaille, using Jean Bouguer's (1753) careful explanation of the method as his starting point, extended the technique to the determination of the apparent angular separation between the Sun and the Moon, although he neglected the effects of the lunar parallax. De Lacaille's major innovation was the simplification of the method from having to understand and execute complex spherical trigonometric calculations to simple geometric operations involving a ruler, a compass and basic operations—all within reach of contemporary sailors. He believed that an accuracy of 4 arcmin in the resulting longitude was attainable (Boistel, 2015).

It took another decade before further developments of importance occurred, this time across the Channel in Britain. On 18 July 1765, the Board of Longitude decided to award Witchell, a watchmaker, £100 for computing, correcting and printing 5000 copies of his *General Tables of Refraction and Parallax*. The Board also decided

> … that Mr Lyons be likewise employed to make the additions to his own tables … and that the said tables, when that is done, be inserted ... that the Astronomer Royal be desired to furnish the Computers thereof with the proper Books of Tables and whatever else may be necessary to assist them in their computations. (Board of Longitude, 1765: 104) while Richard Dunthorne, Lowndean Professor Roger Long's butler at Pembroke College, Cambridge, and a keen amateur astronomer himself (Glyn, 2002), was appointed as 'Comparer of the Ephemerides and Corrector of the Proofs' of the *Nautical Almanac* and the *Astronomical Ephemeris*. (Dunthorne was the sole 'comparer' for the *Almanac*'s first three issues, for 1767–1769, and remained involved until the 1776 issue.)

However, at the Board's meeting on 28 October 1765, it had become apparent that Lyons and Witchell were unavoidably delayed in completing the calculations for the 1767 ephemeris tables. In response, the Board agreed to pay two additional 'computers', John Mapson and William Wales (see Wales, 2015)—who had both already been appointed to execute these tasks for 1768—an additional sum to help. Moreover, the Board decided

> … that £50 be also given to Mr Israel Lyons, as a recompence for another improved method of computing the same, invented by him, which has been already ordered to be printed at the end of the *Ephemerides*. (Board of Longitude, 1765: 128)

Meanwhile, Dunthorne was working on producing his own version of the ephemeris tables. He was well-versed in this endeavour, since he had published his *Practical Astronomy of the Moon: or, new Tables... Exactly constructed from Sir Isaac Newton's Theory, as published by Dr. Gregory in his Astronomy* in 1739. This publication contained astronomical tables based on Newton's theory of 1702, although with some adjustments. Dunthorne had since continued to refine his method. In 1746, he informed the curator of the Woodwardian Museum in Cambridge,

> [a]fter I had compared a good Number of modern Observations made in different Situations of the Moon and of her Orbit in respect of the Sun, with the Newtonian Theory … I proceeded to examine the mean Motion of the Moon, of her Apogee, and Nodes, to see whether they were well represented by the Tables for any considerable Number of Years … (Dunthorne, 1746: 414)

Dunthorne received £50 from the Board of Longitude in 1771 for having contributed a 'rigorous' method[11] to shortening the tedious calculations involved in clearing lunar distances. Dunthorne's improved method was eventually included in the *Nautical Almanac* for 1802. It became popular for shipboard use among the German-speaking diaspora and in Scandinavia until the beginning of twentieth century.

Returning now to the inception of the *Nautical Almanac*, its first edition, for 1767, was completed in 1766 and published in January 1767. It included lunar distance tables, known as 'comparing distances', for the Royal Observatory in Greenwich in three-hour intervals and for a range of reference stars to enable their use at sea. Intermediate positions could be derived by interpolation using a set of equations provided.

In 1767, Witchell was elected a Fellow of the Royal Society, but Lyons was passed over for that same honour. This was despite the *Nautical Almanac*'s adoption of Lyons' newly invented method for clearing lunar distances. Nevertheless, Lyons' achievements were eventually duly recognised by the scientific community, as exemplified by the praise he received from the Swiss astronomer Jean Bernouilli in 1771:

> … le second [Lyons] n'a peut-être pas d'instrumens astronomiques en propre, mais il a si bien mérité de l'Astronomie & de l'Hydrographie par ses tables & méthodes pour les réductions des observations de la Lune, que je suis bien aise de pouvoir vous dire que j'ai fait sa connoissance. (Bernouilli, 1771: 124–125)

Lyons' most important achievement during this time, as head of the small group of computers responsible for preparing Maskelyne's *Nautical Almanac* tables, was the compilation of the *Tables for correcting the apparent distance of the Moon and a Star from the Effects of Refraction and Parallax*. Commonly known as the 'Cambridge Tables' (Shepherd, 1772),[12] they were the result of a monumental effort in calculating lunar distances. A brief introductory section of seven pages, containing instructions for their use, was followed by some 300,000 lunar distance corrections on 1104 pages, with up to 370 corrections per page. These corrections, although impractical for shipboard use, were provided for each degree of lunar distance between 10 and 120 degrees. All possible combinations of solar and lunar altitudes were evaluated, in steps of one degree. Corrections to the apparent lunar distances for the Moon's horizontal parallax of 53 arcmin and the prevailing mean atmospheric refraction were included. Corrections for the ambient temperature and pressure were also given.

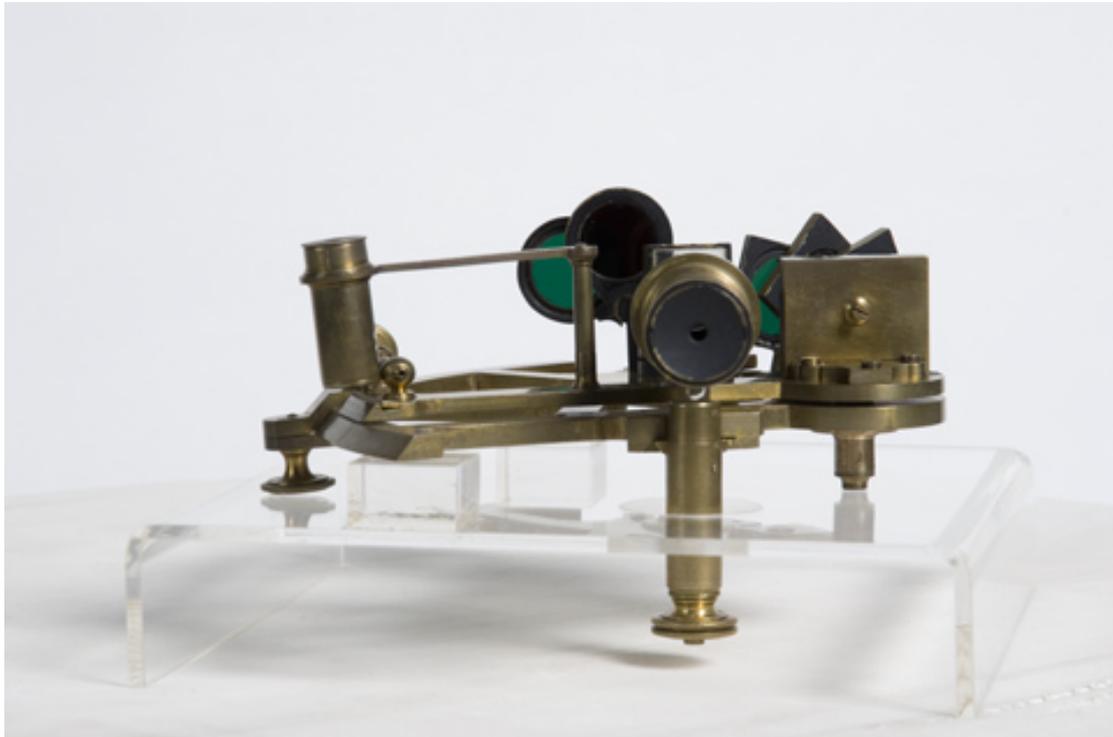

**Figure 7**: James Cook's sextant (Dollond, London, c. 1770). (Courtesy: State Library of New South Wales. Attribution 4.0 International; CC BY 4.0)

In hindsight, 1767 was a crucial year in the development of a viable means of longitude determination at sea. The Board of Longitude finally had a way forward to achieving two potentially practical approaches, the lunar distance method and a sufficiently accurate chronometer. In addition to the rather significant cost of production, manufacture of the latter would take time, however, so that initial efforts centred on application of the lunar distance method.

Captain (First Lieutenant) James Cook was one of the first to adopt lunar distance measurements on the *H.M.B. Endeavour*'s voyage of 1768–1771. At the time, H4 was still the only sufficiently accurate marine chronometer in existence, but it was deemed too precious to be sent on a challenging voyage like Cook's. Cook and his on-board astronomer, Charles Green, applied Maskelyne's *Nautical Almanac* and its associated lunar position tables for 1768 and 1769 (the only tables available at the time of the *Endeavour*'s departure from England), combined with the tried and tested dead reckoning approach. Cook and Green often used a sextant (see Figure 7) to take lunar distance measurements independently but at the same time (Keir, 2010). Their sextant readings often differed by upwards of one arcminute (Wales, 1788: 44, 52), a similar level of uncertainty as that inherent to the lunar table values themselves (Keir, 2010; see also endnote 8; one arcminute is equivalent to one nautical mile at the Equator).

However, even short intervals of inclement weather could disrupt the regular flow of lunar distance observations, and so dead reckoning—combined with a fixed compass course—continued to be routinely used as a fundamental yet supplementary navigation tool. The ship's progress was measured regularly using simple chip floats attached to log lines, combined with a 28 second hourglass as timing device. The ratio of 28 seconds to one hour is the same as that of the rope length between two subsequent knots in the log line (47 feet 3 inches or 14.4 m) to one nautical mile, so this approach enables mariners to obtain a quick measure of their ship's speed, expressed in knots, with respect to the ocean's surface; 1 knot

= 1852 m h$^{-1}$. (Note that in Cook's time the runtime of a standard hourglass was 30 seconds.) However, it did not allow correction for the effects of drift or ocean currents (de Grijs, 2017: 4-40–4-41), and so dead reckoning was only accurate when sailing within coastal view where landmarks could be monitored.

**6 LUNAR DISTANCE TABLES IN MORE MODERN TIMES**

Despite the successful development of a marine timepiece in the late eighteenth century, clearing lunars remained an eminently viable means of longitude determination at sea. During the second half of the eighteenth and the first half of the nineteenth centuries, hundreds of different mathematical approaches were introduced to reduce the burden on marine navigators trying to work out their position at sea by providing pre-calculated lunar distance tables and ever more practical aids. Among those who proposed novel methods or developed new techniques during this period, a few individuals deserve more than a cursory reference to their work. I will discuss their innovations in this section. Nevertheless, and irrespective of a significant uptick in innovative solutions, clearing lunar distances eventually proved too cumbersome compared with the almost direct longitude measurements afforded by the new, increasingly affordable chronometers.

In 1771, the French mathematician, physicist and sailor **Jean-Charles de Borda** embarked on a transatlantic voyage aimed at finding improved methods of longitude determination and testing a number of marine timekeepers. He published his own, 'rigorous' method for longitude determination in 1777 (de Borda *et al*., 1777), which is still regarded as the best of several mathematical procedures in shipboard use prior to the large-scale adoption of accurate maritime timepieces (e.g., van der Werf, 1977). It was particularly popular during the first half of the nineteenth century. His method became a common tool in the arsenal of the French navy, while simplified versions were included in the *Connaissance des Temps* in 1779, 1780 and 1789. Since de Borda's first publication, his method has been subject to frequent further development and improved rigour (e.g., van Swinden, 1789; Chauvenet, 1863).

**János Ferenc (Franz Xaver) von Zach**, a Hungarian-born astronomer, introduced a novel optical arrangement for transit timings, which allowed him to observe his reference stars with his right eye while keeping his left eye on a wall clock. Most important for the current narrative, in 1798 von Zach established his monthly *General Geographical Ephemerides*[13] in Weimar. The publication served as a clearinghouse for observations used in support of geographic longitude determinations. However, after having published just four issues, in 1800 von Zach resigned as the publication's editor; he subsequently established its successor, the *Monthly Correspondence*,[14] which was published until 1813. He continued to actively contribute to fundamental astronomy, however, by publishing books and tables describing the ephemerides of solar system bodies. **Joseph Jérôme Lefrançois de Lalande**, one of France's most respected astronomers, considered von Zach's ephemeris tables the best available at the time (Armitage, 1949).

Between 1772 and 1785, de Lalande himself had laid the foundations for a purely French set of ephemeris calculations based on Maskelyne's tables, aided by his student assistants Edme-Sébastien Jeaurat and Pierre Méchain (Boistel, 2001, 2015). De Lalande had been elected director of the *Connaissance des Temps* in 1759. Nevertheless, the *Minstre de Marine* had only agreed with great reluctance to a request from the *Académie de Marine* to include lunar distance tables, initially translated and taken from the *Nautical Almanac*, in the *Connaissance des Temps*—probably largely driven by an intense cross-Channel rivalry. From 1790, their ephemeris tables were calculated for the Paris meridian, rather than for Greenwich, by the lunar distance computer Louis-Robert Cornelier-Lémery, who was employed full-time for this specific purpose. To their dismay, de Lalande and his team had found a worrying level of mistakes in Maskelyne's tables:

> We are occupied these days in recalculating from observations Maskelyne's 34 stars which we've used with complete confidence, and I find it is necessary to add 5 or 6 seconds to the right ascensions. So we'll have to correct all our catalogues, all our tables and all our longitudes of the observed planets! This old pen pusher, lazy drunkard, miser, has usurped our trust. He's very rich, he should have got himself a computer and checked, more than once, this important result. (Lalande, 1803)

**John William Norie**, mathematician, hydrographer, chart maker and commercial publisher of nautical books, is most famous for his *Complete Set of Nautical Tables* (1803)—commonly known as 'Norie's linear tables', in essence a set of nomograms or graphs[15]—and his *New and Complete Epitome of Practical Navigation* (1805). Norie depended on the sale of his publications for some of his income and so he made sure that they were nearly impossible to copy (original nomogram engravings were very difficult to obtain). Unfortunately, their computational advantage compared with the existing body of lunar tables was fairly trivial, so it appears that their initial sales did not quite meet Norie's expectations. Nevertheless, and probably because of their high accuracy, his tables eventually developed into a standard work on navigation. It included a large number of methods for clearing lunar distances. Norie's tables hence became essential practical tools for British marine navigators in the mid-nineteenth century and they were reprinted numerous times.

**William Lax**, Lowndean Professor of Astronomy and Geometry at the University of Cambridge, served as an astronomer on the Board of Longitude from 1818 until its dissolution in 1828. Although he was tasked with oversight of Maskelyne's *Nautical Almanac*, he also developed his own lunar ephemeris tables. These were first included in the *Almanac* in 1821. Lax's tables were lauded as a "very meritorious attempt to solve the problems of nautical astronomy by one uniform system" by the *Nautical Magazine* (1832: 256; see also Raper, 1840: 897), although they were not very useful in practice (see, e.g., Sabine, 1825: 398). Quite exceptionally for a Commissioner of the Board of Longitude, the Board awarded Lax £1050 for his tables, which were intended to replace Maskelyne's *Requisite Tables* (Waring, 2013). In 1828, he produced an appendix to the *Nautical Almanac*, titled *An easy method of correcting the lunar distance, on account of the spheroidal figure of the earth*.

Having failed to reach Syria for missionary work around the turn of the nineteenth century, **James William Inman** was recommended to the Board of Longitude to join the *H.M.S. Investigator* as astronomer in June 1803, replacing his colleague John Crosley, for exploratory and navigational work in Australian waters (Morgan, 2017). He had trained at the Royal Observatory in Greenwich under the tutelage of the Astronomer Royal. In 1808, he was appointed as Professor of Nautical Mathematics at the Royal Naval College in Portsmouth (England). Inman is best known for *Navigation and Nautical Astronomy for Seamen* (1821), whose tables were used by generations of sailors to solve navigational problems while at sea. Perhaps unusual for the times, he provided a note of caution regarding the reliability of the observations at the basis of his lunar tables:

> The seaman knows very well that caution must be used in trusting too much to Tables of this kind; especially to the longitudes put down, which from the errors of the Lunar Tables at the time the observations for determining then were made, from errors of observations, and other causes, may possibly be wrong a great part of a degree. (Inman, 1821b: 466)

Captain **David Thomson** developed his own, 'short and convenient' (i.e., approximate) methods to clear lunar distances at sea in the early nineteenth century. His lunar tables were first published in 1825. In the December 1824 issue of the *Monthly Critical Gazette*, a reviewer heaped lavish praise on Thomson ("the inventor of the longitude scale") and on the accuracy of his tables:

> … the mode of performing this part of the operation … is extremely simple, and is perhaps the most convenient method of calculation that has ever been offered to the public: it may be thus performed in a third part of the time that has been required for common methods, and this is no small recommendation. (*Monthly Critical Gazette*, 1824)

Thomson's tables became enormously successful commercially in the British merchant navy (*Monthly Critical Gazette*, 1825), to the extent that they were reprinted annually or even more frequently. In response to an unfavourable review of the accuracy of his tables and the ease of their use, Thomson (1827: 146–147) defended himself vigorously and pointed out that

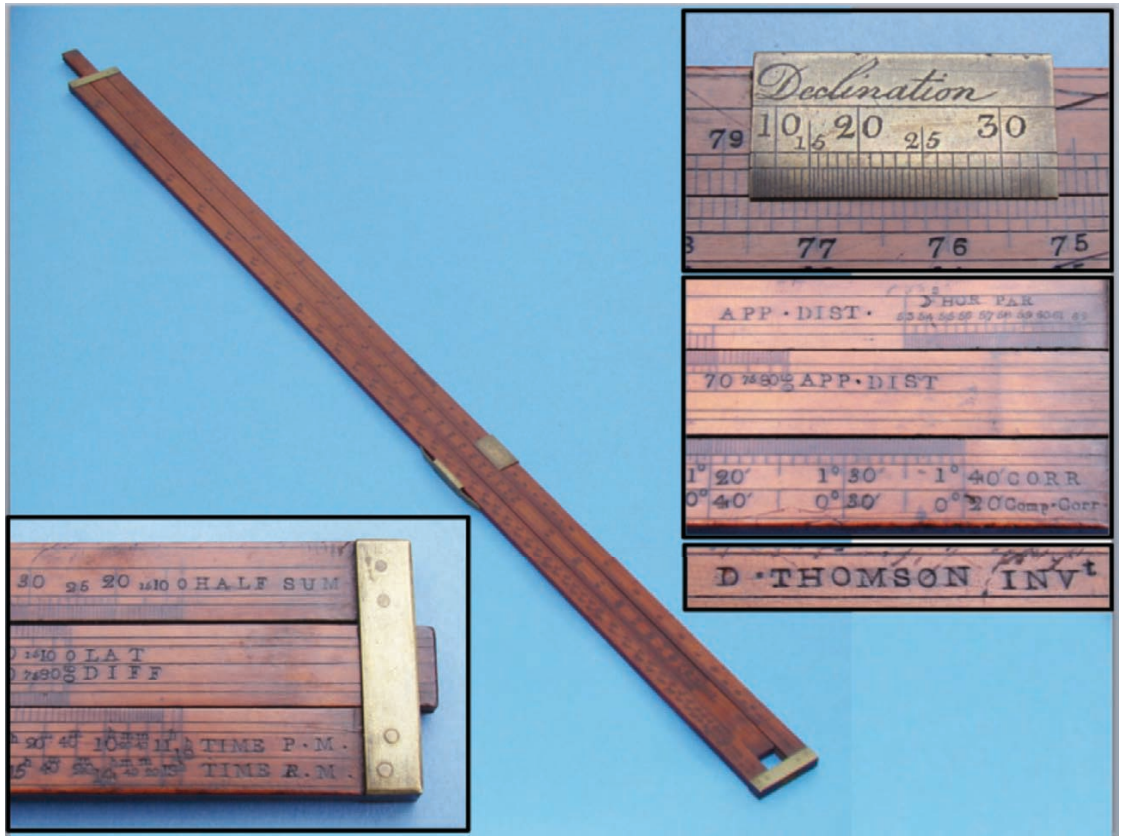

**Figure 8**: A rare Thomson's lunar corrector slide rule, signed 'D. Thomson Invt' and 'Bate London, No 209', made of boxwood, 94 cm long, with brass fittings and cursor. Both sides of the rule and slider are marked with numerous scales (see insets). The brass cursor is engraved with a declination scale. (Courtesy Tesseract – Early Scientific Instruments; reproduced with permission)

> … in none of [my tables] does the true distance, as computed by my method, differ so much as three seconds from the truth; and there are only four examples [provided by his critic], out of the thirty-eight given, where the difference amounts to two seconds [or more].

He then proceeds,

> In the last example given by your correspondent, to show the error to which my method is liable, he makes the error sixteen seconds, when it is only seven and a half seconds.

Thomson's tables contained some obfuscations that did not make the navigator's life easier. Instead, they made it significantly harder though not impossible for anyone attempting reproduction. Once again, we see commercial interests playing a role in this context. He never provided the full details of the corrections applied to his lunar distance measurements. Thomson's tables clearly confused von Zach, leading him to erroneously claim that Thomson must have calculated and interpolated hundreds of thousands of cases exactly by a 'long' method, tabulating the differences. Thomson did no such thing, but von Zach's speculation persuaded others that reproducing Thomson's tables would be too monumental a task. Prospective buyers, on the other hand, seemed awed by the alleged enormity of the work. In addition to his lunar distance tables, Thomson also invented a lunar corrector slide rule to ease the mathematical burden on the navigators of his time (see Figure 8).

Around 1815–1820, numerous authors published lunar tables that would have little or no impact on the scientific enterprise. One such set of tables, known as *Ward's Lunar Tables*, published in North America between the 1820s and the 1850s (e.g., Ward, 1823), caught the eye of von Zach, and they began a life of their own. Secondary mid-nineteenth century sources often include Ward's tables among the most important early tables, despite their unremarkable and rather trivial nature. Interestingly, these tables came to prominence because von Zach discussed them in his third journal, *Correspondence Astronomique* (1818–

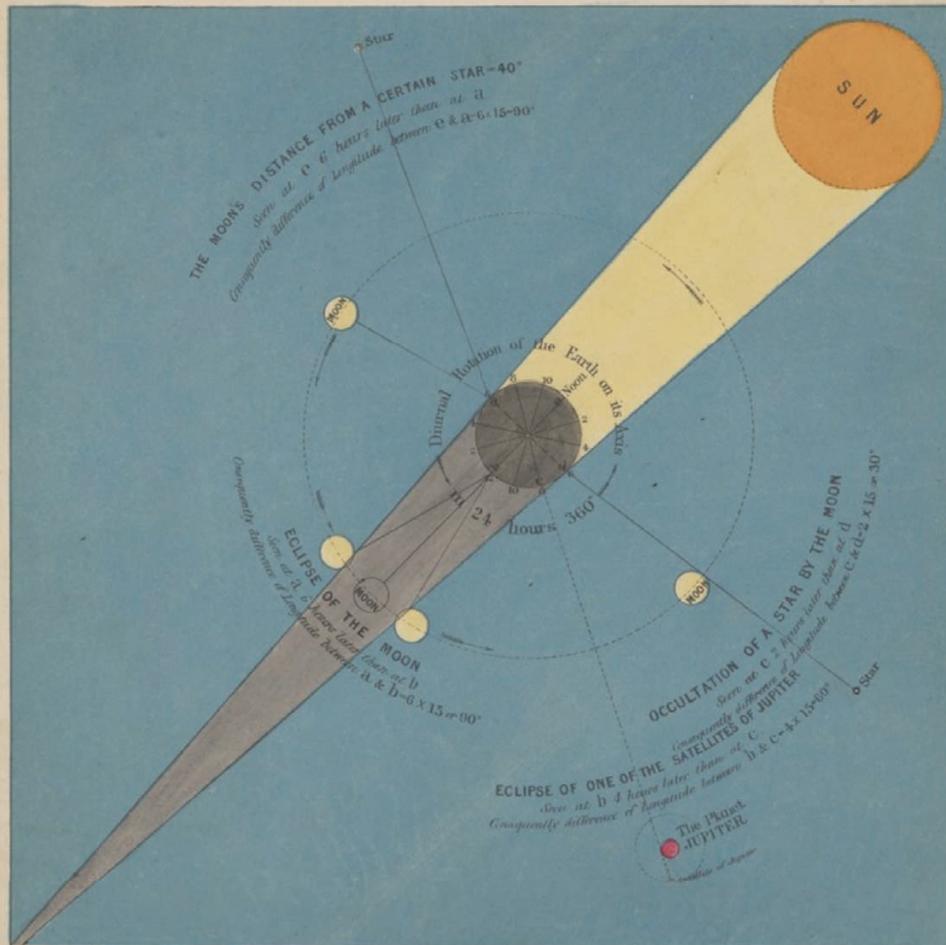

**Figure 9**: Diagram showing how to determine longitude using astronomical observations. Coloured engraving by John Emslie, published by J. Reynolds (1851). (Credit: Wellcome Collection. Creative Commons, Public Domain Mark)

1825). Around 1820, Ward had written von Zach, by then a man of significant means and elevated to minor nobility status, that his tables were being copied extensively without royalty payments. His intent was for von Zach to help him stop this unauthorised behaviour. Instead, Baron von Zach—given his considerable wealth oblivious to market economics—gullibly published Ward's tables in his own journal, for the world to access free of charge…

Finally, but perhaps not surprisingly, after 1822 Charles Babbage designed his programmable mechanical computer in part to help solve the corrections required to clear lunar distances. The design was based on precursors to punched cards. Unfortunately, its prototype did not work. The state of the art of astronomical approaches to longitude and time determinations is summarised well in the 1851 print by John Emslie shown in Figure 9.

**7 THE END OF AN ERA**

Despite the increasing accessibility of ever more practical aids, including most prominently Maskelyne's *Nautical Almanac* and Bowditch's (1802) *New American Practical Navigator*[16] (Thornton, 2006), the lunar distance method eventually proved too cumbersome compared with the simpler use of increasingly more affordable chronometers. Moreover, since the time at sea between ports decreased, chronometer-based time keeping became ever more reliable. This famously led Lecky and Allingham (1918) to declare the lunar distance method "deader than Julius Caesar" by about 1850. Indeed, from the first edition of their handbook, published in 1881, Captain Lecky (1881; cited by van der Werf, 1997) boldly declared that

> … [t]he writer of these pages, during a long experience at sea in all manner of vessels [since the 1850s] …, has not fallen in with a dozen men who had themselves taken Lunars, or even had seen them taken.

(For a remarkable exception, the '*tables du négrier*', which were most likely based on tables published at the beginning of the nineteenth century, see Dubois, 1881.)

Yet, perhaps one of the most famous applications of the lunar distance method occurred as late as 1896, during the solo circumnavigation of the Earth by the American adventurer Captain Joshua Slocum. Slocum departed in his sloop, the *Spray*, from his base at Fairhaven (Massachusetts) on his circumnavigation attempt on 24 April 1895. He owned a chronometer, a remnant of his previous occupation as captain-owner of a moderately sized sailing ship, but it needed costly repairs of about $15. Slocum may have lacked the funds to have it repaired, or perhaps he was simply reluctant to spend this sum of money, arguing that "in our newfangled notions of navigation it is supposed that a mariner cannot find his way without one; and I had myself drifted into this way of thinking" (Slocum, 1900: Ch. II). As a compromise, however, he bought an old tin clock, which had been discounted from $1.50 to $1.00.

Apparently, his tin clock worked well during the initial stages of his voyage. Slocum crossed the Atlantic twice—first to Gibraltar, continuing to South America—before crossing the Strait of Magellan into the Pacific Ocean. It appears that during this time, he resorted to measuring meridian altitudes to set a course by dead reckoning:

> On September 10 [1895] the *Spray* passed the island of St. Antonio, the northwesternmost of the Cape Verdes, close aboard. The landfall was wonderfully true, considering that no observations for longitude had been made, (Slocum, 1900: Ch. V)

although on a later leg, from Rio de Janeiro, Brazil, to Uruguay, we learn that

> … the steamship *South Wales* spoke to the *Spray* and unsolicited gave her the longitude by chronometer as 48° W, "as near as I can make it", the captain said. The *Spray*, with her tin clock, had exactly the same reckoning. (Slocum, 1900: Ch. VI)

However, the old tin clock's reliability seems to have deteriorated by the time of Slocum's Pacific crossing—the only instruments he had access to during that crossing were a sextant and his tin clock. He attempted to regain his time reference using a lunar distance observation. Following his departure from the island of Juan Fernández off the Chilean coast on 5 May 1896,

> … on the forty-third day from land [16 June 1896]—a long time to be at sea alone,—the sky being beautifully clear and the moon being 'in distance' with the sun, I threw up my sextant for sights. I found from the result of three observations, after long wrestling with lunar tables, that her longitude agreed within five miles with that by dead reckoning. (Slocum, 1900: Appendix)

Slocum's account implies that the southernmost of the Marquesas Islands was about to come into view at the time, and he is clearly proud of his achievements:

> The result of these observations naturally tickled my vanity, for I knew that it was something to stand on a great ship's deck and with two assistants take lunar observations approximately near the truth. As one of the poorest of American sailors, I was proud of the little achievement alone on the sloop, even by chance though it may have been. (Slocum, 1900: Ch. XI)

He most likely picked up a copy of the *Nautical Almanac* for 1896 in either Montevideo (Uruguay) or Buenos Aires (Argentina), or possibly already in Gibraltar, for the princely price of 2 shillings and sixpence. He points out that he found an error in the *Almanac*'s lunar tables:

> The first set of sights … put her many hundred miles west of my reckoning by account. I knew that this could not be correct. In about an hour's time I took another set of observations with the utmost care; the mean result of these was about the same as that of the first set. Then I went in search of a discrepancy in the tables, and I found it. In the tables I found that the column of figures from which I had got an important logarithm was in error. It was a matter I could prove beyond a doubt, and it made the difference as already stated. (Slocum, 1900: Ch. XI)

The 'error' is most likely a 12 hour shift in the Moon's right ascension and declination (van der Werf, 1977). Although the *Almanac* listed the times of noon and midnight clearly, the equivalent timings for the Moon's right ascension and declination are confusing. They misleadingly seem to suggest the use of civil rather than astronomical time. At the time of Slocum's circumnavigation, the astronomical day started at the mean Sun's Greenwich meridian passage; the civil day started at midnight, 12 hours earlier.

On the final legs of his voyage, starting with his Indian Ocean crossing, his tin clock lost its minute hand and it had to be boiled periodically to keep it running. Although Slocum appears to have set his course based on longitude determination read off from the Sun's meridian transits, lunar distance measurements must have aided his time determination following the clock's demise.

Publication of lunar tables in the *Nautical Almanac* continued until 1907, while appendices clarifying how to calculate and correct predicted lunar distances were included until 1919 (and until 1924 in the *Nautical Almanac Abridged for the Use of Seamen*; Sadler, 1968). Lunar tables in the *New American Practical Navigator* were updated until 1914. They were no longer included in the francophone *Connaissance du Temps* from 1905 (which was announced by Guyou, 1902). Not everyone agreed with these decisions, as evidenced by a number of heated discussion pieces published in the *Nautical Magazine* between 1900 and 1905. Yet although application of the lunar distance method for longitude determination was gradually superseded by the use of increasingly accurate chronometers, the lunar distance method was still often used to calibrate chronometer performance. Even today, lunar tables are kept up to date by a small number of enthusiasts and specialised organisations (Brunner, 2005; Stark, 2010; Romelczyk, 2019).

> On the forty-third day from land, a long time to be at sea alone, the sky being beautifully clear and the moon being "in distance" with the sun, I threw up my sextant for sights. I found from the result of three observations, after long wrestling with lunar tables, that her longitude by observation agreed within five miles of that by dead reckoning. ... I sailed on with self-reliance unshaken, and with my tin clock fast asleep. … The work of the lunarian, though seldom practised in these days of chronometers, is beautifully edifying, and there is nothing in the realm of navigation that lifts one's heart up more in adoration. (Slocum, 1900: Ch. XI)

## 8 NOTES

[1] Figure 1 is a rogues' gallery of portraits of the main characters driving the developments described in this paper, if and when available in the public domain. I encourage the reader to refer to this figure whenever a new personality is introduced in a more than cursory manner.

[2] I have adopted the convention to refer to Chinese names in the form 'surname followed by first name'.

[3] Shipboard observers were exclusively male.

[4] British Library Add. MS Birch 4393, f. 89; holography copy with signature of the King and Williamson (PRO/SP44/334/27-8).

[5] In calm weather at sea, experienced observers can measure lunar distances to an accuracy of about 0.25 arcmin, resulting in a longitude uncertainty of up to 0.25 degree or 15 nautical miles at the Equator (e.g., Romelczyk, 2019).

[6] By the 1760s, Euler had engaged a team of young assistants, including Anders Johan Lexell, Wolfgang Ludwig Krafft and his own son, Johann Albrecht, to rework the prevailing lunar theory using extensive calculations based on mathematical tables. In the title of the resulting book, Euler praised his team's "incredible zeal as well as the indefatigable labour of three academicians" ("*incredibili studio atque indefesso labore trium academicorum*").

[7] The brightest fixed stars used for lunar distance determination include Aldebaran, Algol, Antares, Arcturus, Bellatrix, Betelgeuse, Canopus, Castor, Deneb, Pollux, Procyon, Regulus, Rigel, Sirius, Spica, Vega and many others.

[8] In the early 1700s, lunar positions could be calculated to an accuracy of about 0.5 acrmin, corresponding to a longitude error of 0.25 degree at the latitude of Greenwich. These uncertainties were roughly halved by 1810. It took until 1860 for the uncertainties to be reduced to 0.1 arcmin, equivalent to the instrumental margin of error at sea. This corresponds to a positional uncertainty at the Equator of around 3 nautical miles (Romelczyk, 2019).

[9] *Tobias Mayer*; https://en.wikipedia.org/wiki/Tobias_Mayer [accessed 23 March 2020].

[10] In 1731, Edmond Halley, Britain's second Astronomer Royal, had published a method based on lunar occultations to correct lunar tables and thus determine longitude at sea (Halley, 1731).

[11] 'Rigorous' methods worked through the full extent of the lunar distance corrections required, including for the Earth's ellipsoidal shape and for the prevailing thermometer and barometer readings. These methods were too cumbersome for everyday use on board ships.

[12] Although Anthony Shepherd contributed the preface to the *Cambridge Tables*, he was not involved in the calculations contained in them.

[13] *Allgemeine geographische Ephemeriden*, Verfasset von einer Gesellschaft Gelehrten, und herausgegeben von F. von Zach, Weimar, 1798, etc.

[14] *Montaliche Correspondenz zur Beförderung der Erd- und Himmelskunde*, herausgegeben von Fr. Von Zach, Gotha, 1800, etc.

[15] One of the first graphical representations was published by James Maud Elford (Charleston, SC), a British merchant navy captain turned American citizen, in 1810. It saw extensive use and 'adoption' by other authors until the end of the nineteenth century.

[16] Bowditch's (1802) *New American Practical Navigator* contained descriptions of four methods of longitude determination through lunar distance measurements. His novel method is almost identical to the method published just a few years earlier by de Mendoza y Ríos (1797, Appendix). It is possible that some cross-fertilisation occurred between the lunar table developments by Bowditch and de Mendoza y Ríos; http://fer3.com/arc/m2.aspx/Lyons-methods-for-clearing-lunar-distance-FrankReed-jun-2013-g24254 [accessed 27 March 2020].

## 9 ACKNOWLEDGEMENTS

I thank Weijia Sun for helping me source the portraits of the Chinese scholars included in Figure 1. I also acknowledge a number of helpful suggestions from the reviewers.